\documentclass[pre,preprint]{revtex4}
\usepackage{amsmath}
\usepackage{multirow}
\usepackage{graphicx}
\begin{document}

\title{The network of stabilizing contacts in proteins studied by coevolutionary data}
\author{Sara Lui}
\affiliation{Department of Physics, Universit\`a degli Studi di Milano, via Celoria 16, 20133 Milano, Italy}

\author{Guido Tiana}
\email{guido.tiana@unimi.it}
\affiliation{Department of Physics, Universit\`a degli Studi di Milano, and INFN, via Celoria 16, 20133 Milano, Italy}
\email{guido.tiana@unimi.it}
\date{\today}

\begin{abstract}
The primary structure of proteins, that is their sequence, represents one of the most abundant set of experimental data concerning biomolecules. The study of correlations in families of co--evolving proteins by means of an inverse Ising--model approach allows to obtain information on their native conformation. Following up on a recent development along this line, we optimize the algorithm to calculate effective energies between the residues, validating the approach both back-calculating interaction energies in a model system, and predicting the free energies associated to mutations in real systems. Making use of these effective energies, we study the networks of interactions which stabilizes the native conformation of some well--studied proteins, showing that it display different properties than the associated contact network.
\end{abstract}
\maketitle

\section{Introduction}

It was recently found that the set of correlated mutations in a family of homologous protein sequences is a very rich source of information about the physical proteins of the members of the family, so rich that it is sufficient to predict their native conformation \cite{Morcos:2011jg,Marks:2011bc}. In a nutshell, since proteins are energetically highly optimized \cite{Bryngelson:1987uu,Shakhnovich:1993uh}, the increase in energy associated with a mutation in one of the sites of the protein is often compensated by the mutation in a neighboring site. This phenomenon gives rise to correlations in the mutation pattern of the protein that, if purified by indirect effects, can be used to identify neighboring residues, and thus the native conformation of the protein.

However, the amount of information contained in the correlations between mutated residues is more abundant than a binary outcome about their spatial proximity. In fact, the treatment developed by Morcos and coworkers  \cite{Morcos:2011jg} in the framework of an Ising model, gives direct access to the interaction energies between residues. Given an alignment of protein sequences, it is possible to calculate an effective potential for residue pairs and study its properties. This potential is effective in the sense that accounts for all the factors that contribute to stabilize the native conformation of the protein, including those of entropic origin such as the hydrophobic effect (strictly speaking, it is a free-energy parametrization). On the other hand, it is not expected to be portable, meaning that when calculated for a protein, it cannot be used for other proteins.

In the present work, we first validate the approach, showing that it is possible to extract reliable energies from mutational correlations. This is done both simulating sequence evolution in a model system and back-calculating the interaction energies, and comparing the predictions of the effective energies calculated for specific proteins with  experimental mutation free energies. To reach a good performance in terms of predictive power, it is necessary to adjust some details of the original algorithm, especially in terms of the {\it a priori} probabilities and in the treatment of gaps in the sequence.

Although there are tools, based on molecular force fields, already available to predict the effect of mutations on the stability of proteins \cite{Gue:2002}, the use of effective energies obtained by coevolution data is particularly simple, computationally fast and thus suitable for high--throughput mutations and for large systems.

As a further application, the effective energies obtained for different proteins are used to study how evolution achieves the stabilization of the native conformation. For this purpose, we investigate the stabilization network of four proteins of various sizes. The graphs which describes contacts between residues in proteins have been widely studied in the framework of network theory  \cite{Vendruscolo:2002vq,Dokholyan:2002et,Greene:2003ve,Atilgan:2004jy,Bagler:2005uy}. However, since proteins are frustrated systems, not all contacts play the same role in stabilizing the native conformation. The availability of effective energies allow to discriminate what contacts result from an attractive interaction, and what are present in spite of a repulsive interaction. We show that the properties of the stabilization network, obtained keeping into account only stabilizing contacts, display properties which are different by the full contact network.

\section{Test of the algorithm on back--calculated energies}

Following the approach developed in ref.  \cite{Morcos:2011jg}, we make the hypothesis that the interaction between the residues in the protein can be written in the form
\begin{equation}
U(\{r_i\},\{\sigma_i\})=\sum_{i<j}e_{ij}(\sigma_i,\sigma_j)\Delta(|r_i-r_j|)+\sum_i h_i(\sigma_i),
\label{eq:u}
\end{equation}
where $\sigma_i$ is the type of residues at position $i$ of the protein, $\Delta(|r_i-r_j|)$ is a contact function which takes the value 1 if residues $i$ and $j$ are close in space (i.e., they contain a pair of heavy atom closer than 0.4\AA) and zero otherwise,  $e_{ij}(\sigma_i,\sigma_j)$ is the interaction energy between residues $\sigma_i$ at position $i$ and $\sigma_j$ at position $j$, and $h_i(\sigma_i)$ is a one-body potential that can act on the residues. Note that the interaction energy depends explicitly not only on the type of residues but also on its position (i.e., the energy associated with the same pair of residue types is different if the pairs occupy different positions in the protein).

Assuming that homologous sequences represent fluctuations in the canonical ensamble for a fixed native conformation $\{r^0_i\}$ \cite{Tiana:2004ba}, one can define the probability
\begin{equation}
p(\sigma_1,\sigma_2,...,\sigma_L)=\frac{1}{Z}\exp\left[ -U(\{r^0_i\},\{\sigma_i\}) \right]
\end{equation}
whose marginals are the empirical frequency counts $f_i(\sigma_i)$ and $f_{ij}(\sigma_i\sigma_j)$ that one can extract from a dataset of $M$ aligned sequences of length $L$, that is
\begin{align}
&\sum_{\{\sigma_k|k\neq i\}}p(\sigma_1,\sigma_2,...,\sigma_L)=f_i(\sigma_i) \nonumber\\
&\sum_{\{\sigma_k|k\neq i,j\}}p(\sigma_1,\sigma_2,...,\sigma_L)=f_{ij}(\sigma_i,\sigma_j).
\end{align}
The empirical frequency counts are obtained by the PFAM database \cite{Punta:2012ko} (see below), considering alignments with at least 1000 sequences.
In a mean--field approximation, one can solve the inverse problem and from the empirical frequency counts, and in particular from the associated correlation matrix
\begin{equation}
C_{ij}(\sigma\sigma')=f_{ij}(\sigma\sigma')-f_i(\sigma)f_j(\sigma'),
\end{equation}
can obtain the interaction energy
\begin{equation}
e_{ij}(\sigma\sigma')=(C^{-1})_{ij}(\sigma\sigma').
\label{eq:e}
\end{equation}
These energies are defined with respect to a reference residue type $\sigma_{ref}$, for which one imposes $e(\sigma,\sigma_{ref})=0$ for each $\sigma$.

The inversion of $C_{ij}(\sigma\sigma')$ usually is not straightforward because of the limited statistics in the counts and because of the presence of gaps in the alignment. To solve this problem, and to account for phylogenetic biases \cite{Weigt:2009ba}, we reweight the empirical frequencies as $\tilde{f_i}(\sigma)\equiv \sum_s \delta(\sigma,\sigma_i^s)/m_s$ and $\tilde{f_{ij}}(\sigma,\sigma')\equiv \sum_s \delta(\sigma,\sigma_i^s)\delta(\sigma',\sigma_j^s)/m_s$, where $m_s$ is the number of sequences with similarity larger than 70\%, and we modify them as
\begin{align}
&f_i(\sigma)=\frac{1}{c}\left[ \tilde{f_i}(\sigma)+x\frac{M_{eff}}{q}+y\frac{\sum_j \tilde{f_j}(\sigma)}{L}+z \tilde{f_i}(\sigma) \right]\nonumber\\
&f_{ij}(\sigma,\sigma')=\frac{1}{c'}\left[ \tilde{f_{ij}}(\sigma,\sigma')+ x\frac{M_{eff}}{q^2}+\right.\nonumber\\
&\left.+ \frac{y}{L^2M_{eff}}\sum_{kl}\tilde{f_k}(\sigma)\tilde{f_l}(\sigma')+\frac{z}{M_{eff}}\tilde{f_i}(\sigma)\tilde{f_j}(\sigma')  \right],   
\label{eq:emp}
\end{align}
where $c$ and $c'$ are normalization factors, $q$ is the number of residue types and $M_{eff}=\sum_s 1/m_s$. The  terms multiplied by $x$, $y$ and $z$ are "pseudocounts", that complement the empirical frequencies including an {\it a priori} probability \cite{Altschul:2009bk} depending, respectively, on the overall fraction of residue types, on the overall fraction of residue types in the specific alignment, and on the overall fraction of residue types in the specific pair of positions. They become important when the statistics for a given element of the correlation function is poor. In the work described in ref.  \cite{Morcos:2011jg} only the first term was considered (i.e., $y=z=0$).

The first test we carry out is to start from a random interaction matrix $e^0(\sigma_i,\sigma_j)$ and the native conformation of a protein, generate with the algorithm described in \cite{Shakhnovich:1993} an alignment of sequences which folds to this native conformation, and back--calculate the interaction energies  $e_{ij}(\sigma_i,\sigma_j)$  from the alignment, using Eq. (\ref{eq:e}). Then, one of the sequences of the alignment is chosen at random as the putative protein of interest, and the values of $e_{ij}(\sigma_i,\sigma_j)$ 
for its native contacts compared with the starting $e^0(\sigma_i,\sigma_j)$. The matrix $e^0(\sigma_i,\sigma_j)$ has zero average and unitary standard deviation, the interaction energy of glycine is set to zero as reference, the external fields $h$ are set to zero and the selective temperature of the algorithm used to generate the sequences is $T_s=0.5$. This procedure is repeated  20 times and the average correlation coefficient $r$ is used as a measure of the quality of the prediction. 

\begin{figure}
\includegraphics[width=8cm]{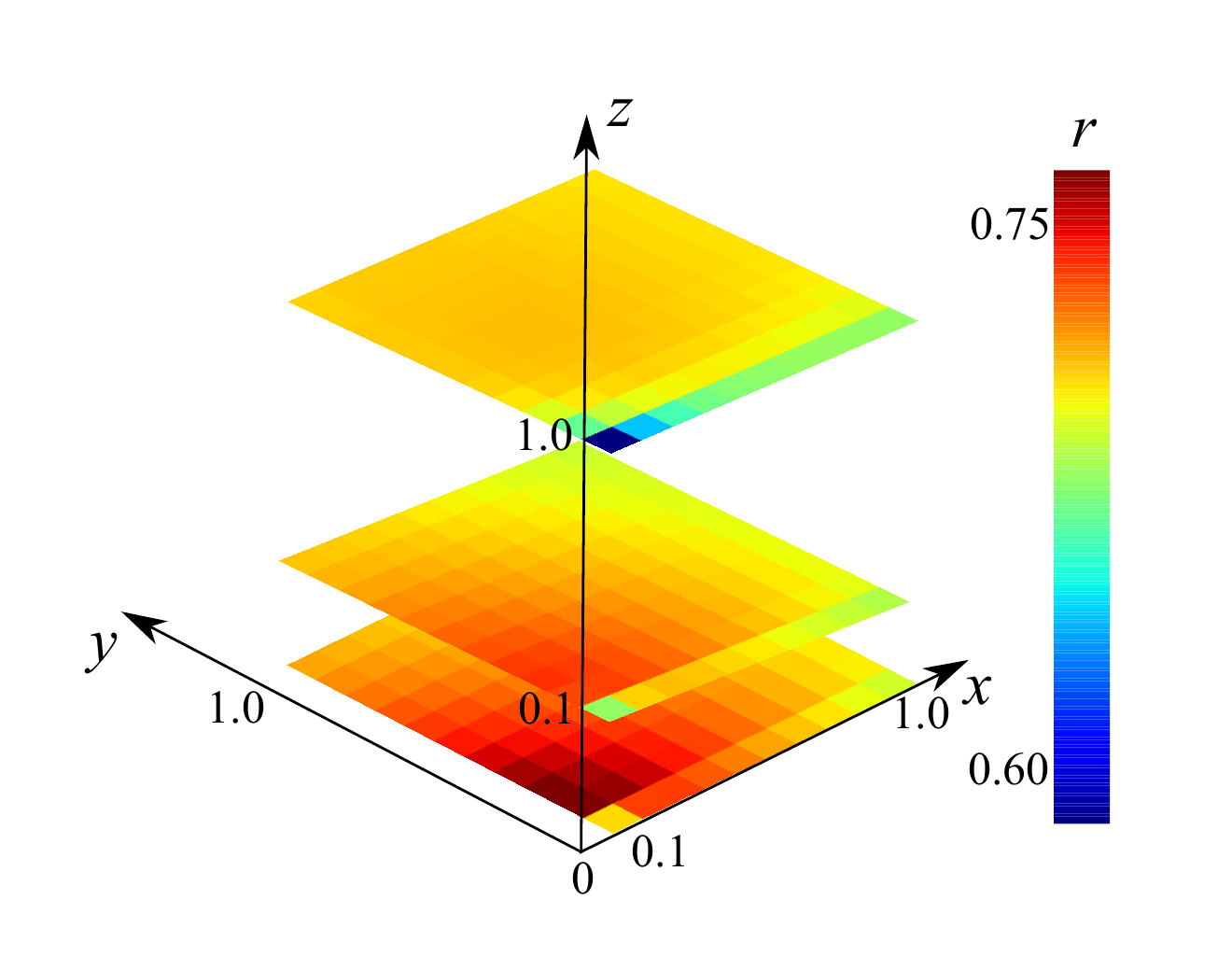}
\caption{The correlation coefficient $r_{model}$ between model and back--calculated energies in absence of gaps, as a function of the pseudocounts $x$, $y$ and $z$.}
\label{fig:nogap}
\end{figure}

In Fig. \ref{fig:nogap} we plot the value of $r_{model}$ as a function of $x$, $y$ and $z$ for ACBP, a protein of 86 residues \cite{Kragelund:1999il}. The best result found is $r=0.76$ for $x=0.1$, $y=0.1$ and $z=0$. This means that the {\it a priori} probability accounting for the probability of residues type in a specific site does not improve the prediction of the energy, but that associated with the specific alignment does. In fact, the value of $r_{model}$ obtained with the choice $x=0.7$, $y=0$ and $z=0$ used in \cite{Marks:2011bc} gives $r=0.70$, whole the optimisation of $x$ only allows for an improvement limited to $r_{model}=0.73$.

In order to check the dependence of the results on the specific protein, we also tested the case of the SH3 domain of Src \cite{Grantcharova:1998vw} and the PDZ domain of Ptp-BL \cite{Gianni:2007js}, obtaining for the best choice of the parameters described above $r_{model}=0.71$ and $r_{model}=0.79$, respectively.

\section{Treatment of the gaps in the alignment}
 
Real alignments usually are different than the model case described above because of the presence of gaps, due to insertions and deletions in the evolutionary dynamics of the family of homologous proteins. We have tested several methods of treating gaps: \\
(M1) A gap is considered as a type of residues, thus $q=21$.  The reference residue $\sigma_{ref}$ is one of the amino acids. This is the choice done in ref.  \cite{Morcos:2011jg}. \\
(M2) A gap is considered as a type of residues, thus $q=21$, but at variance with (M1) sequence with more than 30\% gaps are  discarded, as described in ref. \cite{Rivoire:2013}. \\
(M3) Same as M1, but choosing the gap as reference type $\sigma_{ref}$. \\
(M4) Same as M2, but choosing the gap as reference type $\sigma_{ref}$. \\
(M5) In each site, each gap is considered as a different kind of residues, so that $q$ is site--dependent and is $20+g_i$, where  $g_i$ is the number of gaps in site $i$. \\
(M6) A gap is considered as a type of residues, so $q=21$.  Empirical frequencies in the two Eqs. (\ref{eq:emp}) are weighted by $(1-g_i/M)$ and $(1-[g_i+g_j]/2M)$, respectively, to give more weight to sites with less gaps. To compensate for this effect, it is added the {\it a priori} probabilities $g_i M_{eff}/Mq$ and $(M_{eff}[g_i+g_j]/2q^2M)$, respectively.\\

\begin{figure}
\includegraphics[width=8cm]{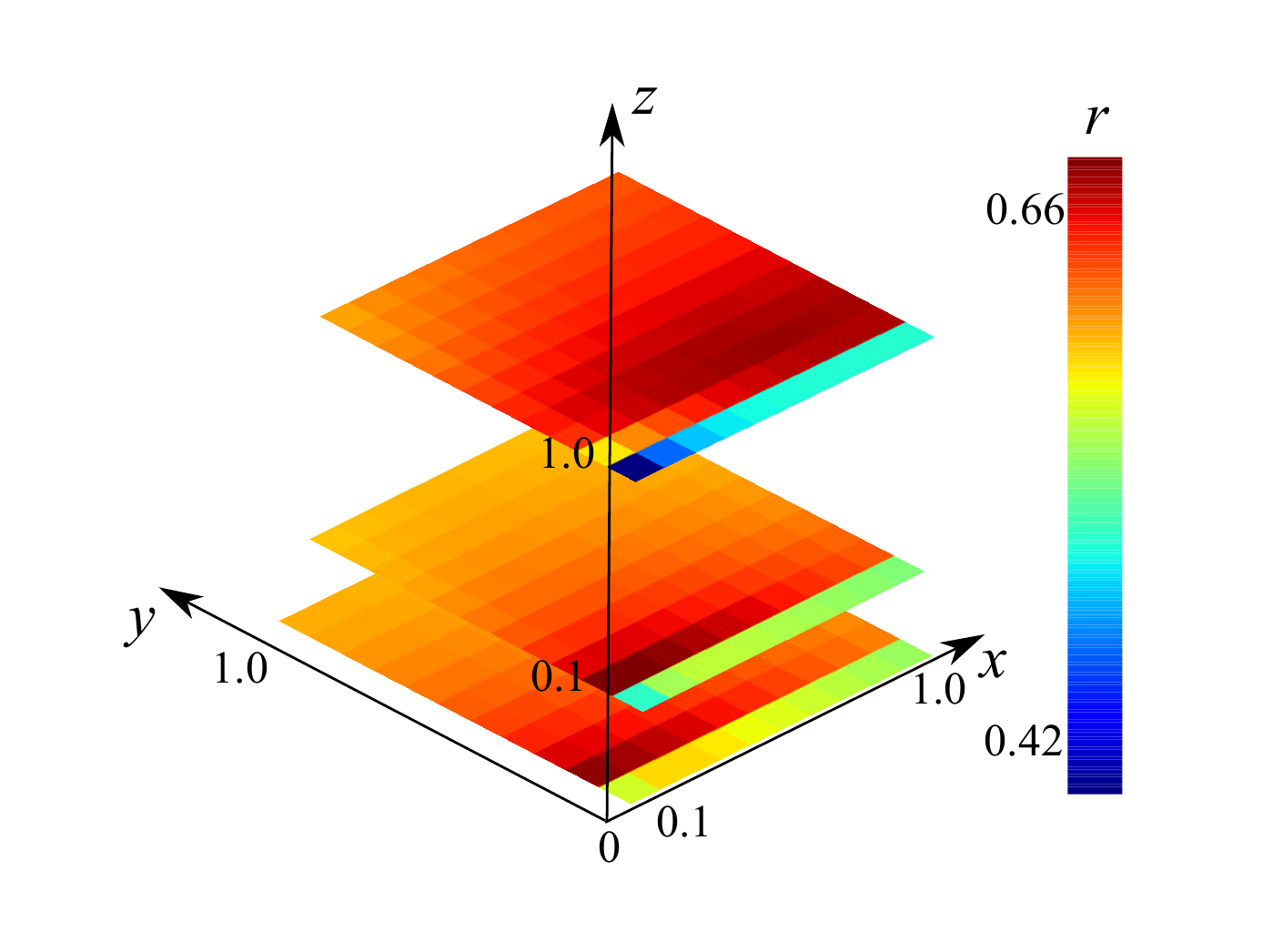}
\caption{The correlation coefficient $r$ between model and back--calculated energies for ACBP sequences with 41 gaps, treated with the algorithm M4, as a function of the pseudocounts $x$, $y$ and $z$.}
\label{fig:gap}
\end{figure}

In order to find the best among the above methods, we have generated model alignments of sequences interacting with a random matrix $e^0(\sigma_i,\sigma_j)$, and then back--calculated the interaction energies making use of Eq. (\ref{eq:e}). At variance with what described in the previous section, here we introduce gaps in the sequences, modelling them as sites of the protein which do not interact with the others. The average correlation coefficient between the energies $e^0(\sigma_i,\sigma_j)$ and their back--calculated values are listed in Table \ref{tab:r} in the case of few gaps (2 out of 86 sites) and of many gaps (41 out of 86 sites). In the case of few gaps, using methods M3 and M4, that is chosing as $\sigma_{ref}$ the gap, and optimizing $x$, $y$ and $z$ gives results comparable with those obtained without gaps. In the more realistic case of many gaps, 
the best results are obtained with methods M3, M4 and M6. In the case of M3 and M4, the highest correlation is obtained without the site--specific {\it a priori} probability (i.e, $z=0$), although unlike the case without gap, choices of $z>0$ do not worsen the results dramatically (cf. Fig. \ref{fig:gap}).

To further validate the approach, and to discriminate among M3, M4 and M6, the present approach has been challenged to reproduce the energetic effect of mutations in known proteins. We have studied four proteins, that is ACBP \cite{Kragelund:1999il},  the PDZ domain of Ptp-BL \cite{Gianni:2007js}, zinc--substituted azurin (AZR) \cite{Wilson:2005} and the SH3 domain of Src \cite{Grantcharova:1998vw}, in which the destabiling free energy $\Delta\Delta G^{exp}$ upon mutations of several residues into alanines have been measured. For each of these proteins an alignment with its homologs has been extracted by the Pfam database (cf. Table \ref{tab:prot}), and the interaction energies  $e_{ij}(\sigma_i,\sigma_j)$ have been calculated with each of the methods described above. Noticeably, this is a 4--dimensional tensor that contains the interaction energies between each pair of sites in contact in the protein, for any pair of amino acids that can occupy those sites. The elements $e_{ij}(\sigma_i,\sigma_j)$ are effective energies, which take into account also an entropic contribution. Assuming, as usually done when interpreting the results of mutation experiments \cite{Fersht:2002vx}, that the interaction energy in the denatured state is negligible, one can easily calculate the energy difference upon the mutation $\sigma(i)\to\sigma'(i)$ as
\begin{equation}
\Delta\Delta G^{calc}\equiv \sum_j [e_{ij}(\sigma'_i,\sigma_j) - e_{ij}(\sigma_i,\sigma_j)]\Delta(|r_i-r_j|).
\label{eq:ddg}
\end{equation}

\begin{figure}
\includegraphics[width=8cm]{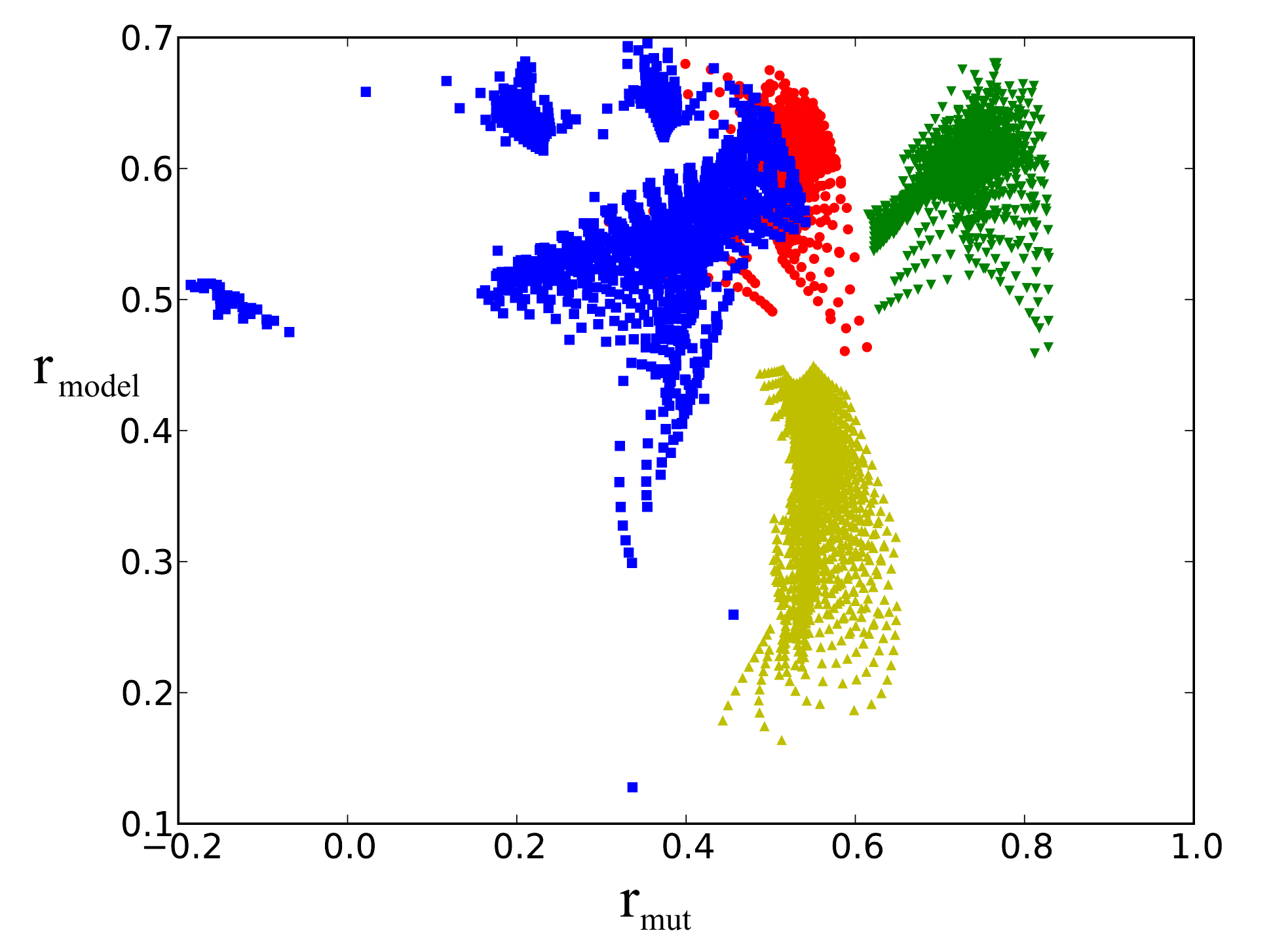}
\caption{For some of the algorithms used in the caluclations it is shown the correlation coefficient $r_{model}$ between model and back--calculated energies and the correlation coefficient $r_{mut}$ between calculated and experimental values of $\Delta\Delta G$ upon mutations. Yellow points are associated with the different values of $x$, $y$, $z$ using M1, red points with M3, green points with M4 and blue points with M6.}
\label{fig:model_mut}
\end{figure}

The best value of the correlation coefficients $r_{mut}$ between calculated and experimental values of  $\Delta\Delta G$, averaged over the four proteins under study, is $r_{mut}=0.82$, obtained with the method M4 and the choice $x=1$, $y=2$ and $z=10$. This is essentially the same of the average correlation reached with FOLD--X \cite{Gue:2002}, an algorithm specifically designed to caluclate mutational free energies, based on a molecular force--field and involving the conformational optimization of the mutants, which gives $r_{mut}=0.83$.   As shown in the scatter plot of Fig. \ref{fig:model_mut}, this is not a particularly good choice from the point of view of the back--calculation of energies in the model discussed above (it corresponds to $r_{model}=0.53$). Since we are interested in an algorithm (i.e., a choice of the method and of $x$, $y$, $z$) that can both back--calculate correctly the interaction energies in the model and reproduce the experimental mutational free energies, we focus our attention on the points that lie in the upper--right region of Fig. \ref{fig:model_mut}. This identified certainly method M4 as the best trade-off between the two requirements, and the choice $x=0.2$, $y=0.1$, $z=0.1$ seems also reasonable, which gives $r_{mut}=0.76$ and $r_{model}=0.68$. This is the algorithm that is used in the following calculations.

\begin{figure}
\includegraphics[width=8cm]{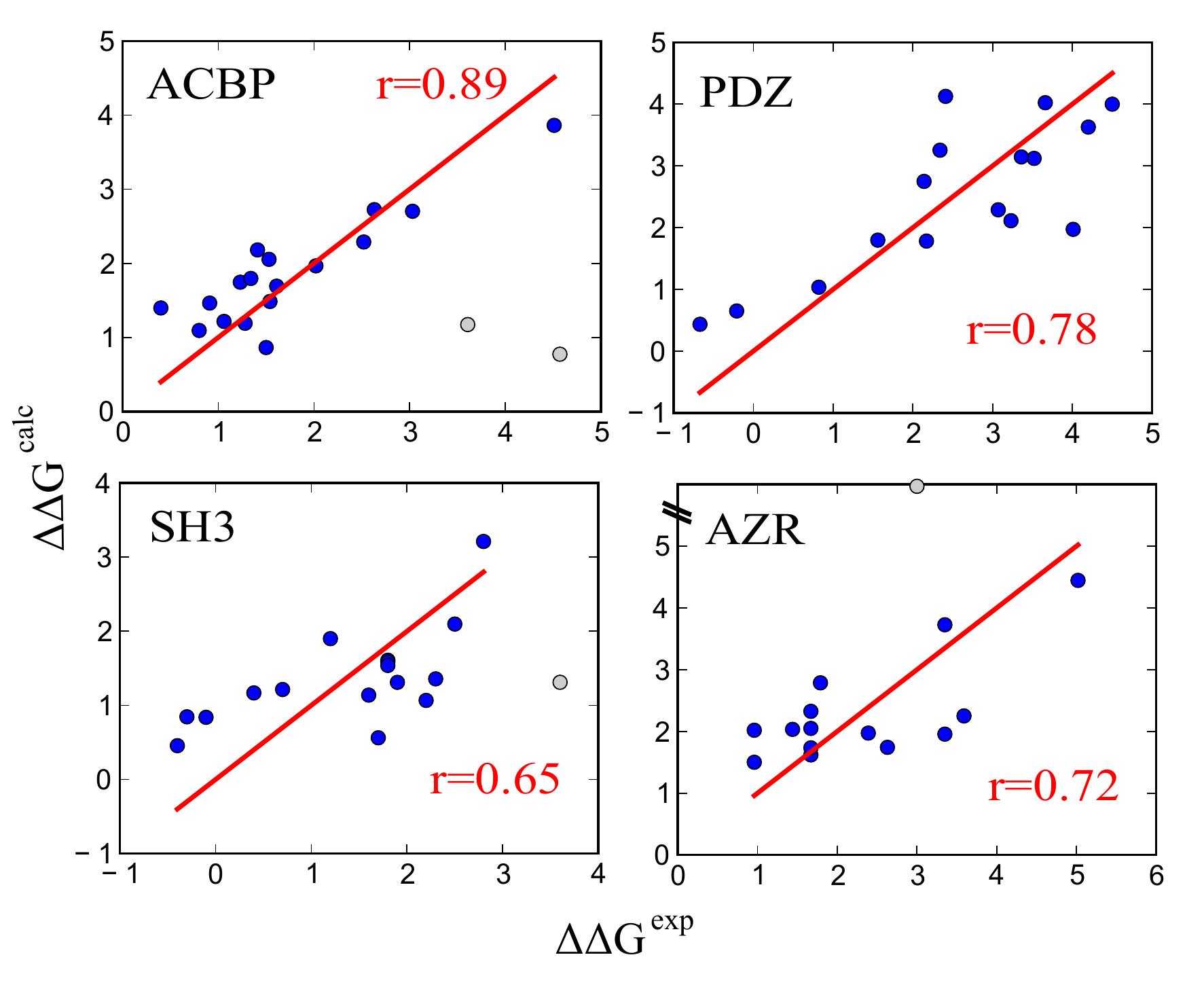}
\caption{The effect of mutations into alanine of the stability $\Delta\Delta G^{exp}$ measured experimentally compared with that calculated by the model $\Delta\Delta G^{calc}$, for ACBP, PDZ domain, src--SH3 and azurine (AZR). The correlation coefficient $r$ is indicated for each protein. Outliers are marked in grey.}
\label{fig:mut}
\end{figure}

The best results for the four proteins, obtained with method M4, are displayed in Fig. \ref{fig:mut}. The correlation coefficients between calculated and experimental data range between 0.65 and 0.89. This comparison shows some outliers, namely mutations L80A and L15A in ACBP, W42A in SH3 and M121A in azurin. In the first two cases, mutations involve residues which have been shown to be structured already in the denatured state \cite{Fieber:2004ww,Rosner:2010kg}, invalidating Eq. (\ref{eq:ddg}). In the case of azurin a characterization of the denatured state is not available. However, the value of $\Delta\Delta G$ for mutation M121A in azurin is site 121 is two orders of magnitude larger than typical values, due to the exceedingly strong interaction between H117 and M121. This is an artifact due to the fact that M121 and H117 are conserved in than 95\% of the alignment of azurin, making the statistics used to calculate the correlations in the mutations of site 121 with the other sites of the protein too poor to allow a reliable estimate of the interaction energy. Consequently, although the qualitative fact that the interaction between M121 and H117 is strongly attractive is realistic, causing such a strong conservation of the two residues, the quantitative estimate of the interaction energy is not.

\section{Quantification of frustration in proteins}

A further advantage of the present treatment of gaps is that, setting the overall zero in residue interaction, provides the information necessary to assess if pairs of residues are attracting or repelling each other on the typical time scale of protein motion.  Statistical potentials, obtained from the count of neighboring amino acids in globular proteins, can provide effective interaction energies (also accounting for entropic effects), but they are defined but for an additive and a multiplicative constant \cite{Tiana:2004ba}. In real proteins, that can fluctuate to non-compact conformations, the energy zero between two residues  is defined when they are far away from each other. That of setting to zero the interaction with a gap is a physically-sound choice that allows to assess which pairs of residues attract and which repeal each other in absolute terms (i.e., with respect to an elongated conformation). In this respect, gaps in the alignment are not an obstacle in the determination of the interaction potential between  amino acids, but a tool to set their zero.

Consequently, it is now possible to investigate which amino acids repel each other in a protein (i.e., which amino acids would be, locally, in a more favourable situation exchanging their actual neighbours with a gap). Since the interaction between amino acids is complex, proteins are likely to be frustrated systems, that is systems that cannot get rid of unfavourable interactions even in the lowest--energy states \cite{Van:1977}. Theoretical arguments suggest that natural proteins are those in which evolution has minimized the degree of frustration \cite{Bryngelson:1987uu,Shakhnovich:1993uh}. The knowledge of interaction energies allows to quantify such a frustration. 

In Fig. \ref{fig:histoe} it is shown the distribution of interaction energies which stabilize the four proteins discussed above.  The fraction of repuslsive energies is 7.5\% in ACBP, 14.3\% in PDZ,  9.8\% in SH3, and 16.7\% in azurin, indicating that evolution cannot eliminate, in average, one repulsive contact every ten.

\section{Properties of the stabilization networks}

The proximity of residues in the native conformation of proteins forms networks which have been widely studied in the past \cite{Vendruscolo:2002vq,Dokholyan:2002et,Greene:2003ve,Atilgan:2004jy,Bagler:2005uy}. These network are, as a rule, of small-world kind, displaying a small average path length $L$ (that is, the average distance between any pair of nodes, in units of number of links) and a large clustering coefficient $C$ (that is, the average fraction of pairs of neighbors that are also neighbors of each other). The small-world character is a consequence of the presence of a small number of residues through which most paths in the network have to go through, as quantified by their "betweeness" $B_i$ \cite{Vendruscolo:2002vq}.

The knowledge of the interaction energies between residues allow to refine the study of the network, selecting only of interactions which contribute sizingly to the stabilization of the protein. From the proximity network, in which each residue is a node and two nodes are linked if the distance between any pair of atoms in the two residues is closer than 4\AA, we create a stabilization network, keeping only the links between pairs of residues whose interaction energy is lower than the threshold $e_t$. In building the network we did not discard interactions of residues close along the sequence, since the interaction between their side chains are also important in stabilizing the protein, and there is no reason to think that the present algorithm treat them worse than the others. Moreover, discarding them has strong effects on the resulting network \cite{Greene:2003ve}.

\begin{figure}
\includegraphics[width=8cm]{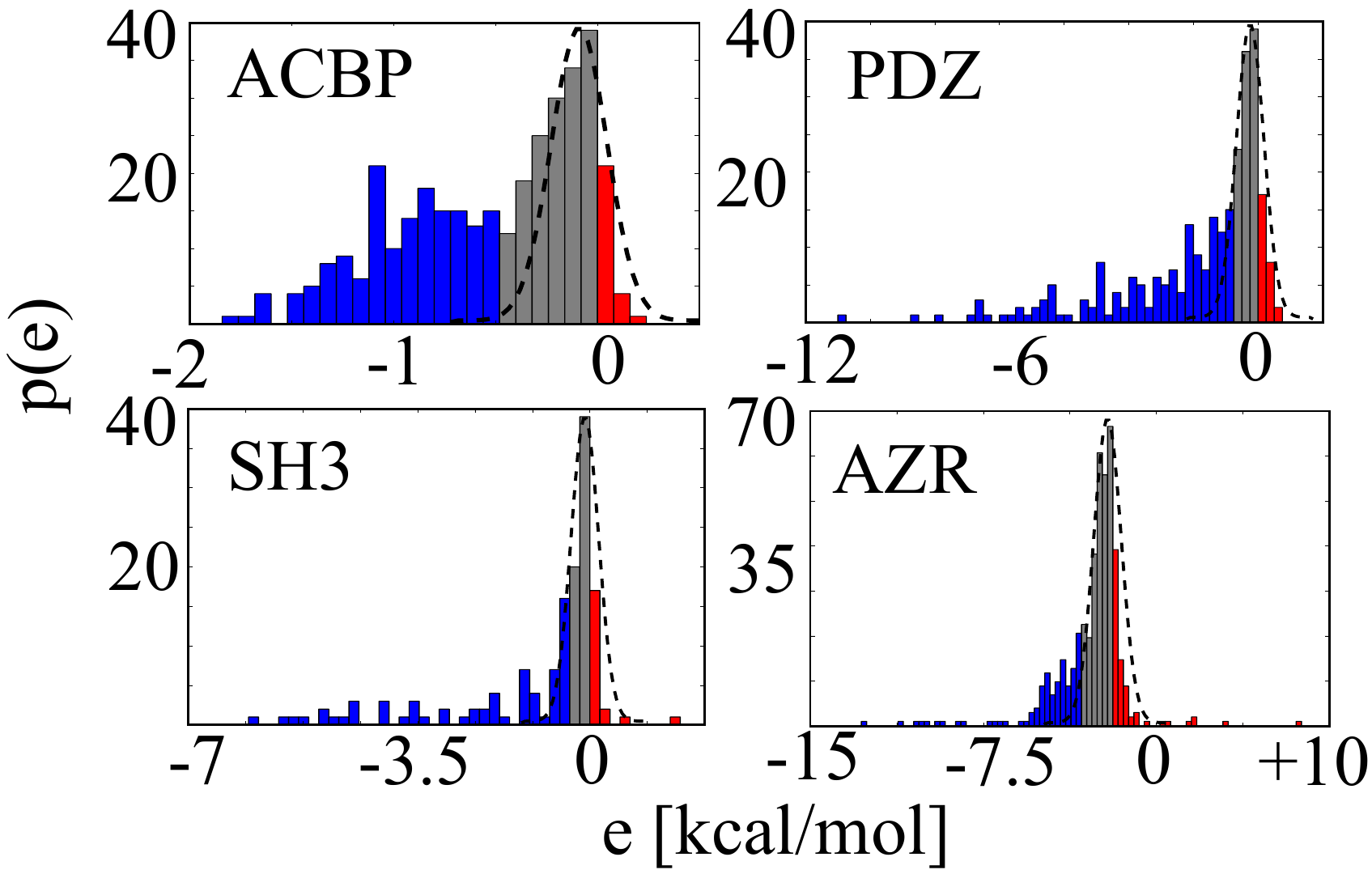}
\caption{The distribution of interaction energies between residues in ACBP, PDZ domain, SH3 and azurin (AZR). The conversion from internal energy units to kcal/mol is based on the comparison with experimental mutational data. The dashed curve is a Gaussian meant to help visualazing the peaked part of the distribution around zero energy. The colors help to distinguish between positive energies (in red), negative energies above the threshold $e_t^*$ (in grey) and below $e_t^*$ (in blue).}
\label{fig:histoe}
\end{figure}

\begin{figure}
\includegraphics[width=8cm]{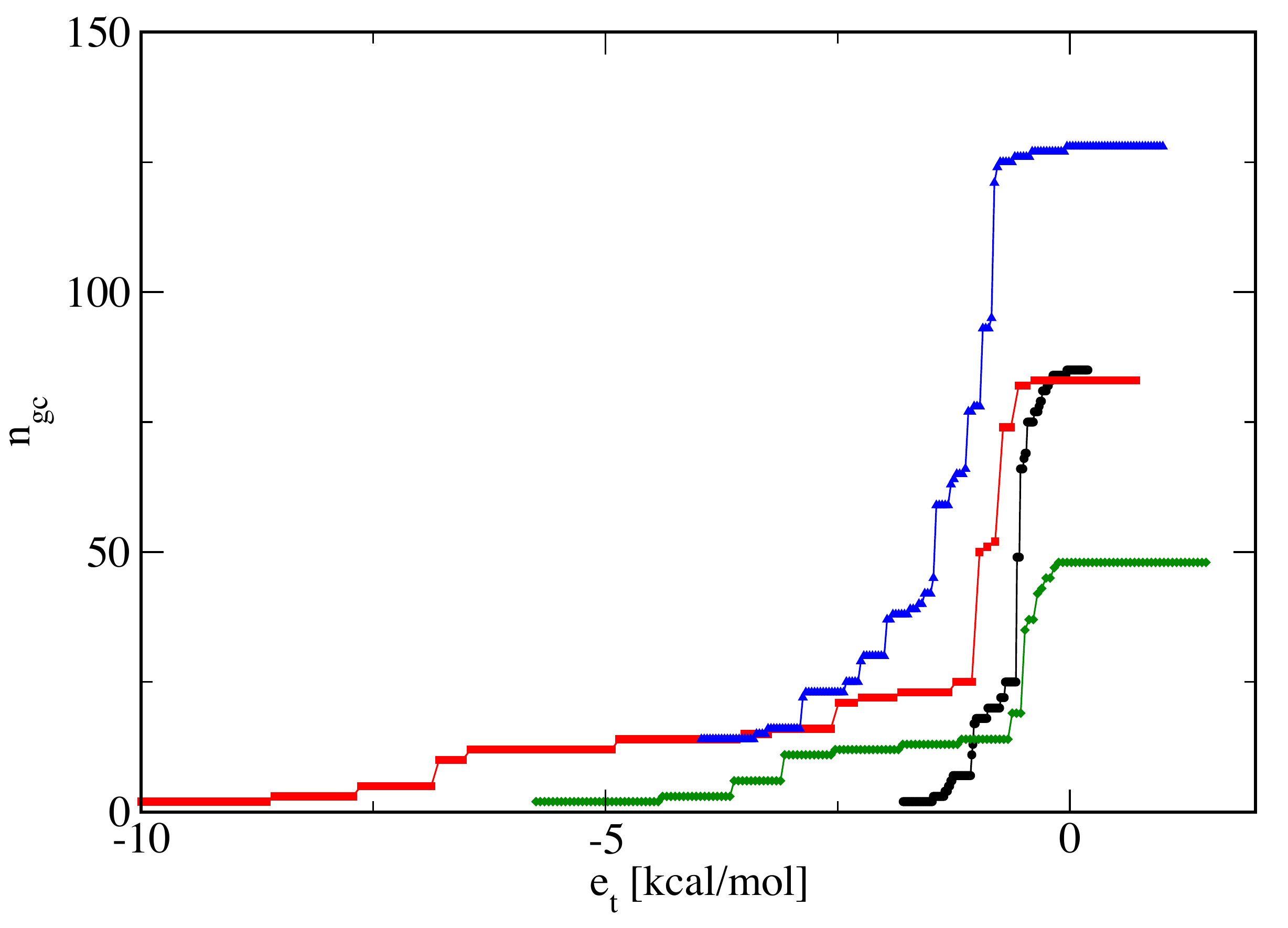}
\caption{The size of the giant component of the stabilization network as a function of the energy threshold $e_t$ used to defined a link for ACBP (black symbols), PDZ (red symbols), SH3 (green symbols) and azurin (green symbols).}
\label{fig:gc}
\end{figure}

To choose a meaningful value of $e_t$, we display in Fig. \ref{fig:gc} the size of the giant component of the network as a function of $e_t$ for the four proteins discussed above. In all cases the size of the giant component displays a sharp jump (at $e_t=-0.48$ kcal/mol for ACBP, $-0.76$ kcal/mol for PDZ,  $-0.34$ for SH3), separating the trivial case in which most of residues are unlinked "orphans", to the case of a single fully--connected network. We will denote this threshold as $e_t^*$. Interestingly, the values of $e_t^*$ correspond to a change of shape in the energy distributions $p(e)$ displayed in Fig. \ref{fig:histoe}. In fact, the $p(e)$ of the four proteins display a Gaussian--like peak centered at zero energy (cf. dashed curve in Fig. \ref{fig:histoe}) and a low--energy tail. This suggests that the links in the stabilization network can be divided into two qualitatively--different groups, namely the weakly--interacting links which populate the Gaussian--like peak, and the highly--optimized contact which populate the tail. The fact that the boundary between the two groups is exactly $e_t^*$ indicates that the highly--optimized links, which are rougly half of the total, are sufficient to maintain the integrity of the network.

\begin{figure}
\includegraphics[width=8cm]{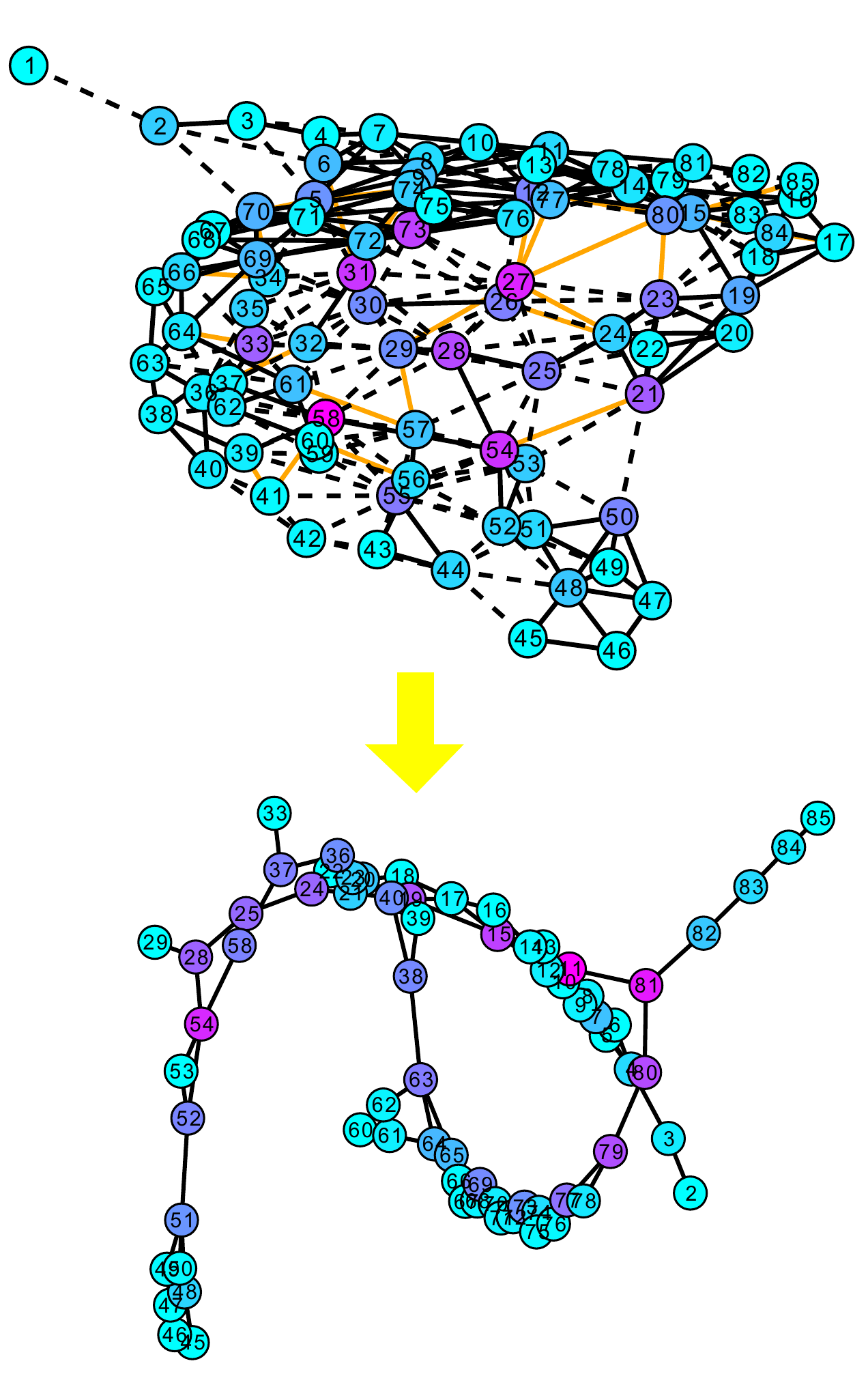}
\caption{Above, the contact network of ACBP. The color of each node, labelled with the residue number, indicate the betweeness $B_i$ (light--blue means low, red means high). Solid links indicate interactions with $e<e_t^*$, dashed links interactions with $e_t^*<e<0$ and orange links interactions with $e>0$. Below, the stabilization network, resulting from keepinf only links with $e<e_t^*$. The color of the node indicates the betweeness of the stabilization network.}
\label{fig:net_acbp}
\end{figure}

\begin{figure}
\includegraphics[width=8cm]{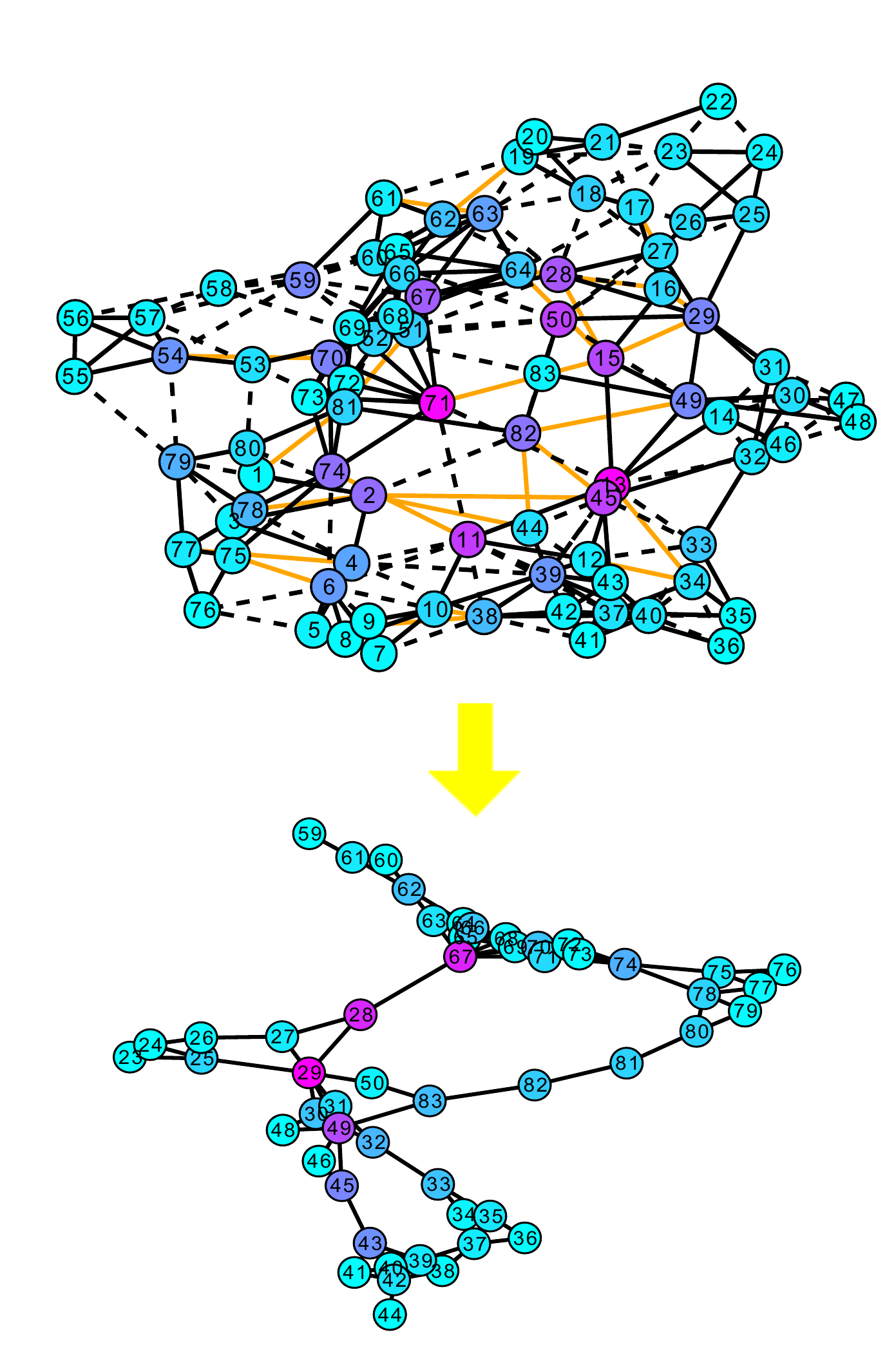}
\caption{Same as Fig. \protect\ref{fig:gc}, for PDZ.}
\label{fig:net_pdz}
\end{figure}

\begin{figure}
\includegraphics[width=8cm]{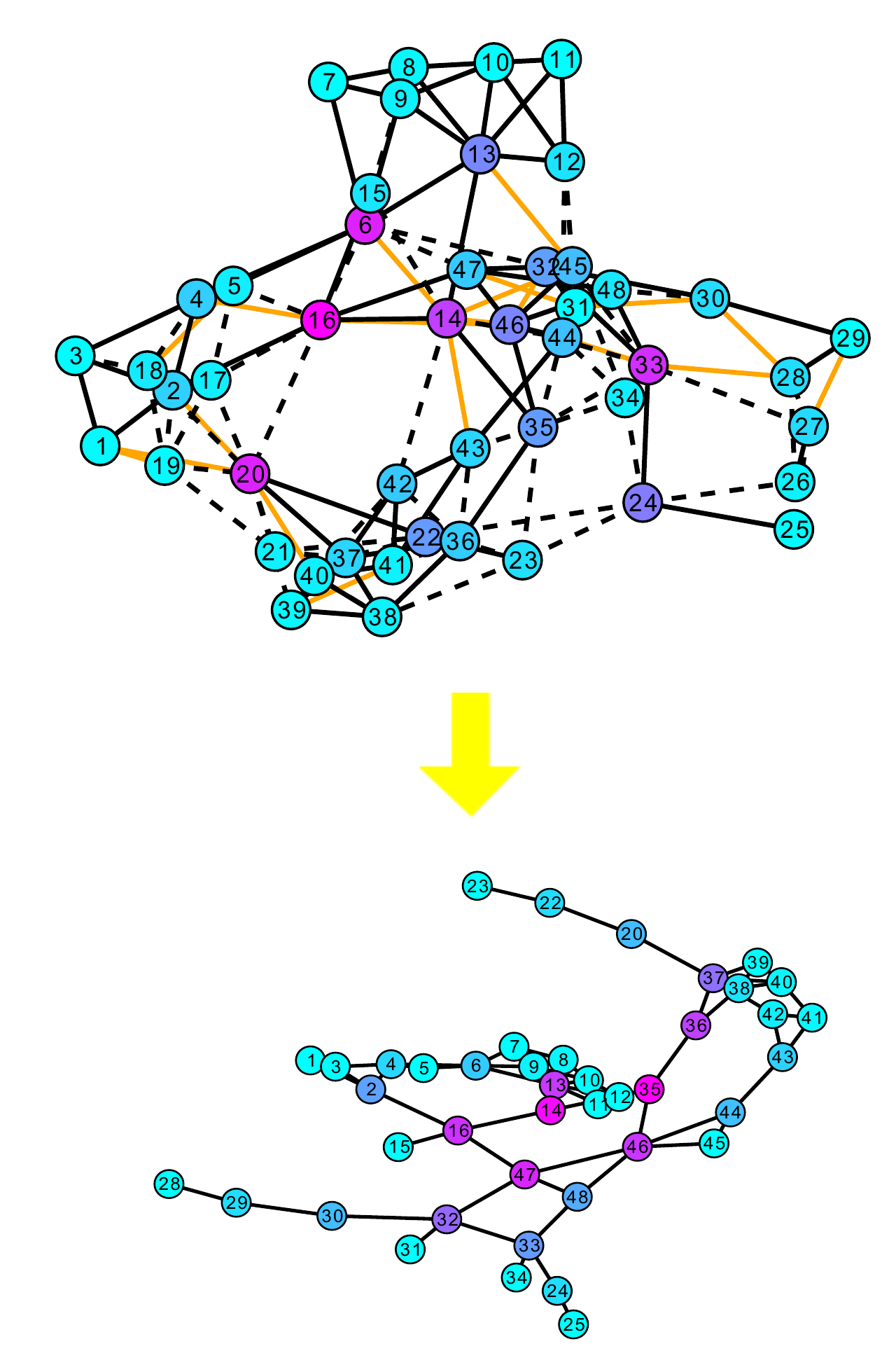}
\caption{Same as Fig. \protect\ref{fig:gc}, for SH3.}
\label{fig:net_sh3}
\end{figure}

\begin{figure}
\includegraphics[width=8cm]{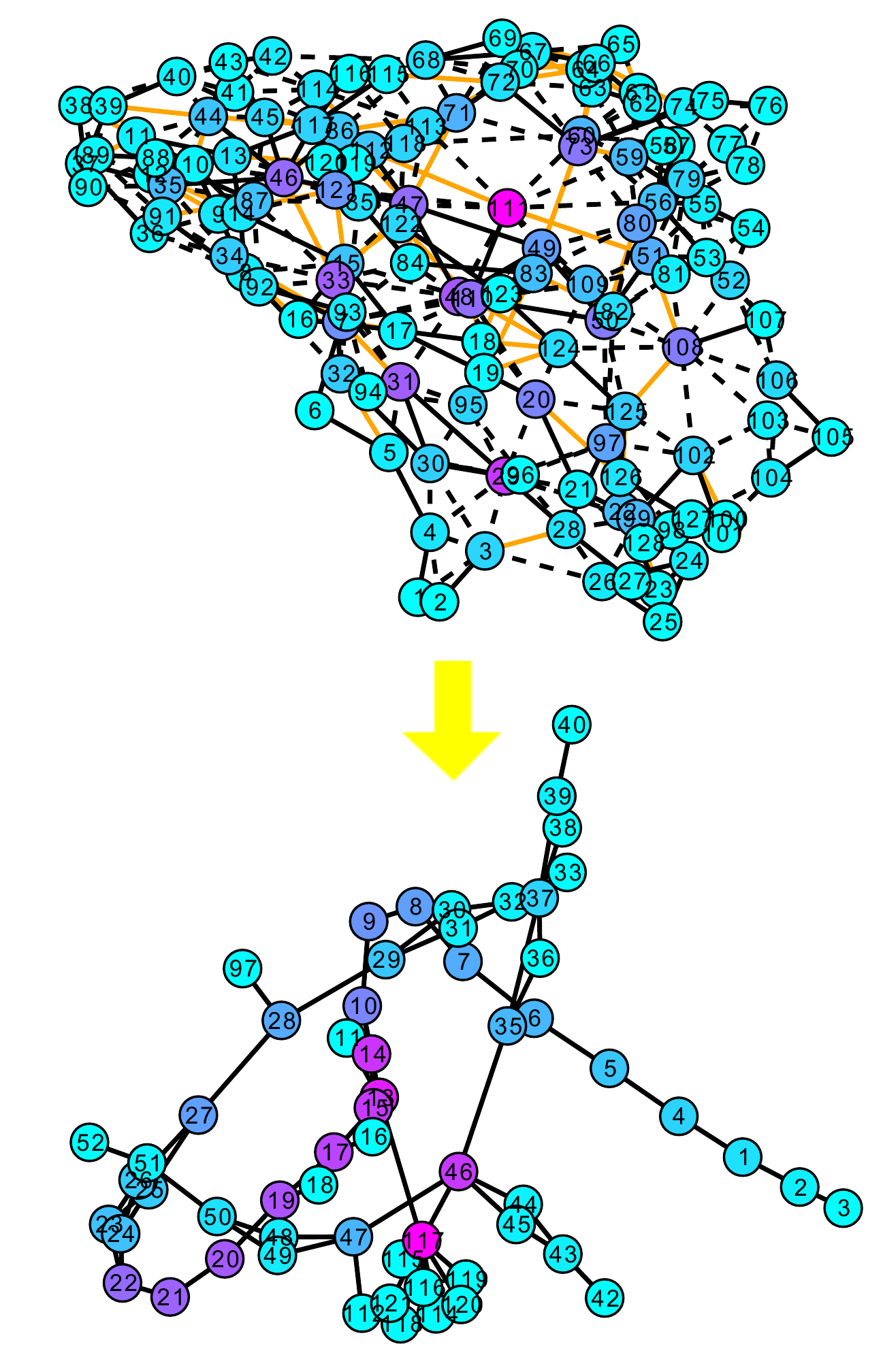}
\caption{Same as Fig. \protect\ref{fig:gc}, for azurin.}
\label{fig:net_azr}
\end{figure}

In order to study the distribution of energies at a molecular level, we plot in Figs. \ref{fig:net_acbp}, \ref{fig:net_pdz}, \ref{fig:net_sh3}, \ref{fig:net_azr} the contact networks (above) and the stabilization networks (below), obtained keeping only interactions energies lower than $e_t^*$. The betweeness of each node is indicated by its color. Both the contact network (as already noted in ref.  \cite{Vendruscolo:2002vq}) and the stabilization network display few nodes displayng high betweeness, while most of them display small betweeness. This behaviour is that typical of scale--free network, although the degree of the nodes can vary in such a small range, due to steric constrains, that the analysis of the distribution of nodes is not feasible here. Moreover, it agrees with the presence of few key sites, critical for folding, found in the study of minimal--model proteins \cite{Tiana:1998td}
% -> red links

Interestingly, nodes displaying large betweeness are very different in the geomteric and in the stabilization network of each protein. For example, in the case of ACBP residues 27, 31, 54, 58 are the most central in the contact network, while residues 11, 15, 54 and 81 are central in the stabilization network. The correlation coefficient between the betweeness of geomteric and of the stabilization network is 0.20 in ACBP, 0.43 in PDZ, 0.59 in SH3 and 0.10 in azurin. Consequently, the stabilization network predicts "key" residues which are different from those predicted by the contact network. Incidentally, this results questions the applicability of structure--based models in predicting the relative stability of different parts of proteins \cite{Sutto:2006}.

Also repulsive interactions are concentrated in few nodes (cf. orange links in the contact networks of Figs.  \ref{fig:net_acbp}--\ref{fig:net_azr}). In all cases they irradiate from nodes with highest betweeness. Usually these nodes are crossed by strongly interacting links (those which also build out the stabilization network), with the interesting exception of site 27 in ACBP, which is crossed only by repulsive and weak interactions.

Stabilization networks have a much simpler topology than the contact network of the same protein. A common motif is the presence of quasi--monodimensional branches of residues which are close along the protein chain (i.e., $i$-$(i+1)$, $i$-$(i+2)$). The physical consequence is that the interaction between the side chains are expected to make the protein chain more rigid in these segments. This effect involves both segments which have been shown to be stable even in non--native conformations and segments which are not, like segment 65--85 and 45--51 of ACBP, respectively  \cite{Fieber:2004ww}. 

\begin{figure}
\includegraphics[width=8cm]{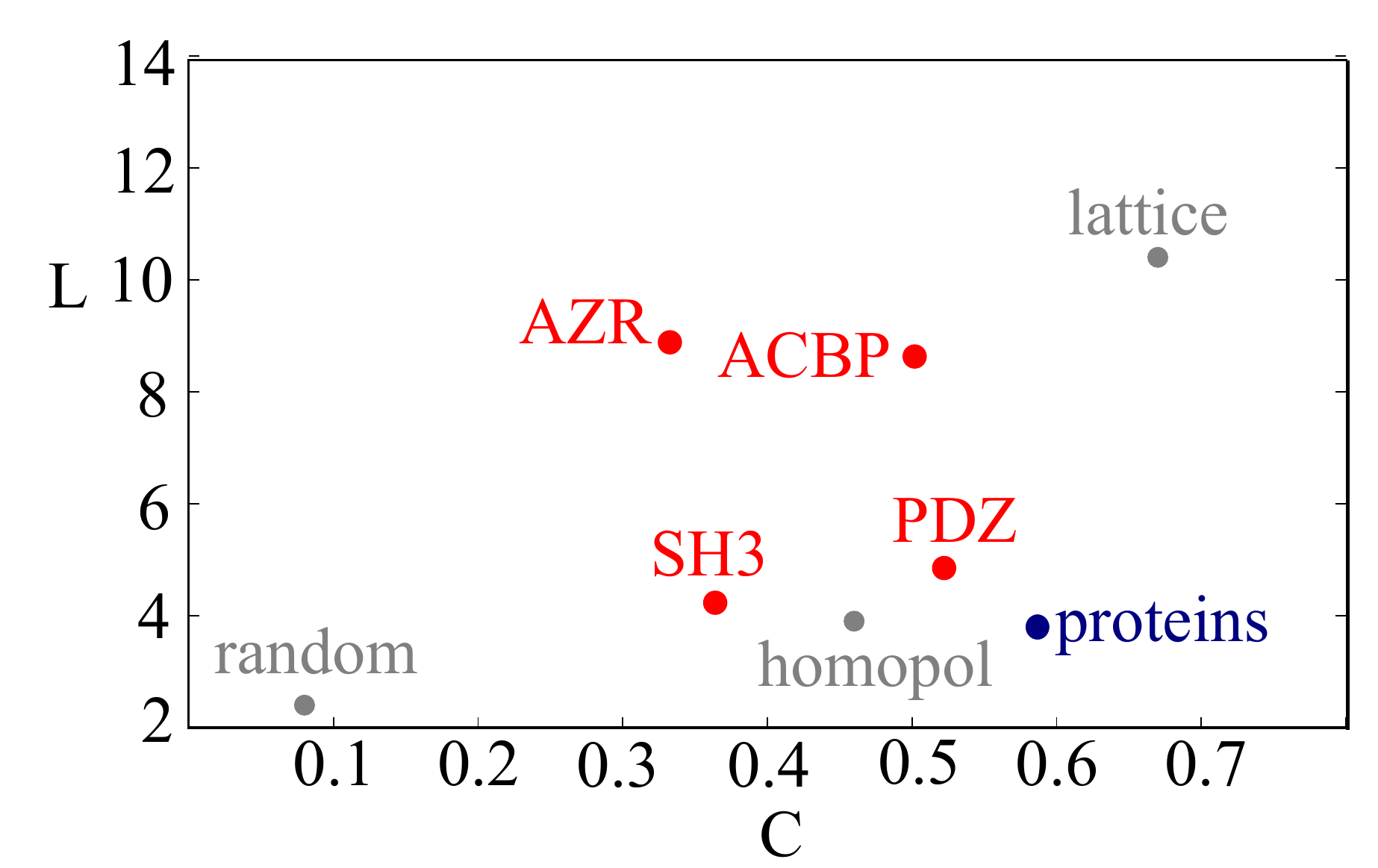}
\caption{The path length $L$ agaist the clustering coefficient $C$ for the stabilizng network of proteins ACBP, PDZ,  SH3 and azurin (AZR). As reference, it is plotted the average value of the same quantities calculated for the proximity networks of proteins calculated in \protect\cite{Vendruscolo:2002vq} (labelled as "proteins" in the plot), for random networks, for homopolymers and for a regular lattice. }
\label{fig:cl}
\end{figure}

The overall connectivity properties of the stabilization network can be summarized in terms of the diameter $L$ and of their clustering coefficient $C$, which are displayed in Fig. \ref{fig:cl} together with the average values of contact networks in proteins, of random networks, of homopolymers and of regular lattices, as calculated in ref. \cite{Vendruscolo:2002vq}. The stabilization network of the four proteins discussed above show a smaller clustering coefficient than that of contact networks of proteins, but still much larger than that of random networks. The diameter is more variable among stabilization networks, but is anyway larger than that of contact networks. Summing up, stabiization networks display less small--world character than stabilization networks, being placed halfway between regular lattices and random networks. In this respect, it fits the definition of aperiodic crystal introduced by Schr\"odinger to describe chromosomes \cite{Schroedinger}.

\section{Conclusions}

Coevolutionary data, obtained by aligning homologous sequences, are one of the most exhaustive and easilly accessible kind of experimental data about most known proteins. The Pfam database contains, up to date, 15.9 million sequences, divided into 13672 families. Consequently, it is important to develop theoretical and computational tools to exploit this richness of information as much as possible.

Following up on the seminal work described in ref. \cite{Morcos:2011jg}, we have developed a strategy to extract effective residue--residue interaction energies from the set of coevolutionary data of a given protein. These energies are not portable, but can anyway be obtained easily for most proteins of interest. One can make use of them to calculate the change of free energy upon mutations, as we did to contribute to the validation of the method. For this purpose, they are particularly simple and computationally handy, and consequently suitable for massive studies or to investigate large proteins. Another use is that of studying the interaction network that stabilizes the native conformation of proteins, with the goal of getting information also concerning the folding process, without any critical constrain on the size of the protein. The stabilization network results into properties which are different than the contact network, like different central residues and a smaller small--word character.

But the possible applications are much more than this. As compared to  force--fields commonly used in modelcular--dynamics simulations,  energies obtained by coevolutionary data have the advantage of being effective energies, also including the effects of entropic origin, like the hydrophobic force. Moreover, they have a well-defined zero, so one can easily understand if two residues attract or repell each other. One can thus use them, for example, to refine simplified molecular models, or to analyze trajectories generated by explicit--solvent molecular dynamics.

%%%%%%%%%%%

\newpage

\begin{table}
\begin{tabular}{|c|c|c|c|c|c|} \hline
 & method & $x$ & $y$ & $z$ &$r_{model}$  \\ \hline
\multirow{2}{*}{no gap} & optimized & 0.1 & 0.1 & 0 & 0.76  \\ 
& parameters of ref. [2] & 0.7 & 0 & 0 & 0.70  \\ 
& optimized with $y=z=0$ & 0.2 & 0 & 0 & 0.73 \\
\hline
\multirow{1}{*}{2 gaps} & M2 optimized & 11 & 25 & 10 & 0.43 \\
 & M3 optimized & 0.1 & 0.2 & 0 & 0.74  \\ 
\hline
\multirow{2}{*}{41 gaps}  & M1 optimized & 15 & 45 & 1 & 0.51  \\ 
& M1, param. of ref. [2] & 0.7 & 0 & 0 & 0.22  \\ 
& M2 optimized & 21 & 43 & 10 & 0.55  \\
& M2, param of ref. [19] & 0.8 & 0 & 0 &  0.24\\ 
& M3 optimized & 0.1 & 0.1 & 0 & 0.68  \\ 
& M4 optimized & 0.1 & 0.1 & 0 & 0.68  \\ 
& M5 optimized & 0.1 & 0.8 & 0 & 0.57 \\ 
& M6 optimized & 0.1 & 0.2 & 0.1 & 0.70  \\ 
\hline
\end{tabular}
\caption{The correlation coefficients between the contact energies of the model and those back--calculated according to different ways to treat the gaps, as described in the text.}
\label{tab:r}
\end{table}

\begin{table}
\begin{tabular}{|c|c|c|c|c|c|}
\hline protein & pdb & length & Pfam id & \# seq. & \% gap\\ 
\hline ACBP & 2ABD & 85 & PF00887 & 1677 & 5.4\\ 
\hline SH3 & 1FMK  & 48 & PF00018 & 10749 & 6.5\\ 
\hline PDZ & 1GM1  & 83 & PF00595 & 26099 & 9.4\\ 
\hline AZR &  5AZU &   128 &  PF00127 & 1417 & 8.9 \\ 
\hline
\end{tabular} 
\caption{The four alignments used in the calculations}
\label{tab:prot}
\end{table}

\begin{thebibliography}{99}
\bibitem{Morcos:2011jg} F. Morcos, A. Pagnani, B. Lunta, A. Bertolino, D. S. Marks, C. Sander, R. Zecchina, J. N. Onuchic, T. Hwa, and M. Weigt, Proc. Natl. Acad. Sci USA {\bf 108}, E1293 (2011)
\bibitem{Marks:2011bc} D. S. Marks, L. J. Colwell, R. Sheridan, T. A. Hopf, A. Pagnani, R. Zecchina, C. Sander, Plos One {\bf 6}, e28776 (2011)
\bibitem{Bryngelson:1987uu} J. D. Bryngelson and P. G. Wolynes, Proc. Natl. Acad. Sci. USA {\bf 84}, 7524 (1987)
\bibitem{Shakhnovich:1993uh} E. I. Shakhnovich and A. Gutin, Proc. Natl. Acad. Sci. USA {\bf 90}, 7195 (1993)
\bibitem{Gue:2002} R. Guerois, J. E. Nielsen, and L. Serrano, J. Mol. Biol. {\bf 320}, 369 (2002)
\bibitem{Vendruscolo:2002vq} M. Vendruscolo, N. V. Dokholyan, E. Paci, and M. Karplus, Phys. Rev. E {\bf 65}, 061910 (2002)
\bibitem{Dokholyan:2002et} N. Dokholyan, L. Li, F. Ding and E. Shakhnovich, Proc. Natl. Acad. Sci USA {\bf 86}, 8637 (2002)
\bibitem{Greene:2003ve} L. H. Greene and V. A. Higman, J. Mol. Biol {\bf 334}, 781 (2003)
\bibitem{Atilgan:2004jy} A. R. Atilgan, P. Akan and C. Baysal, Biophys. J. {\bf 86}, 85 (2004)
\bibitem{Bagler:2005uy} G Bagler and S Sinha, Physica A {\bf 346}, 27 (2005)
\bibitem{Tiana:2004ba} G. Tiana, M. Colombo, D. Provasi, R. A. Broglia, J. Phys. Cond. Mat. {\bf 16}, 2551 (2004)
\bibitem{Punta:2012ko} M. Punta, P. C. Coggill, R. Y. Eberhardt, J. Mistry, J. Tate, C. Boursnell, N. Pang, K. Forslund, G. Ceric, J. Clements, A. Heger, L. Holm, E. L. L. Sonnhammer, S. R. Eddy, A. Bateman, and R. D. Finn, Nucl. Acid Res. {\bf 40}, D290 (2012)
\bibitem{Weigt:2009ba} M. Weigt, R. A. White, H. Szurmant, J. A. Hoch ad T. Hwa, Proc. Natl. Acad. Sci USA {\bf 106}, 67 (2009)
\bibitem{Altschul:2009bk} S. F. Altschul, E. M. Gertz, R. Agarawala, A. A. Sch\"affer and Y.--K. Yu, Nucl. Acid. Res. {\bf 37}, 815 (2009)
\bibitem{Shakhnovich:1993} E. I. Shakhnovich and A. Gutin, Prot. Engin. {\bf 6}, 793 (1993)
\bibitem{Kragelund:1999il} B. B. Kragelund, P. Osmark, T. B. Neergaard, J. Schi¿dt, K. Kristiansen, J. Knudsen, and F. M. Poulsen, Nature Struct. Biol. {\bf 6}, 594 (1999)
\bibitem{Grantcharova:1998vw} V. Grantcharova, D. Riddle, J. Santiago, and D. Baker, Nature Struct. Biol. {\bf 5}, 714 (1998)
\bibitem{Gianni:2007js} S. Gianni, C. D. Geierhaas, N. Calosci, P. Jemth, G. W. Vuister, C. Travaglini-Allocatelli, M. Vendruscolo, and M. Brunori,  Proc. Natl. Acad. Sci. USA {\bf 104}, 128 (2007)
\bibitem{Rivoire:2013} O. Rivoire, Phys. Rev. Lett. {\bf 110}, 178102 (2013)
\bibitem{Wilson:2005} C. J. Wilson and P. Wittung--Stafshede, Biochemisty {\bf 44}, 10054 (2005)
\bibitem{Fersht:2002vx} A. Fersht, {\it Structure and Mechanism in Protein Science}, W. H. Freeman and Co. (2002)
\bibitem{Fieber:2004ww} W. Fieber, S. Kristjansdottir, and F. Poulsen, J. Mol. Biol. {\bf 339}, 1191 (2004)
\bibitem{Rosner:2010kg} H. I. R\"osner and F. M. Poulsen, Biochemistry {\bf 49}, 3246 (2010)
\bibitem{Van:1977} J Vannimenus and G Toulouse, J. Phys. C {\bf 10}, L537 (1977)
\bibitem{Tiana:1998td} G. Tiana, R. A. Broglia, H. E.Roman, and E. Vigezzi and E. Shakhnovich, J. Chem. Phys. {\bf 108}, 757 (1998)
\bibitem{Sutto:2006} L. Sutto, G. Tiana and R. A. Broglia, Protein Sci. {\bf 15}, 1638 (2006)
\bibitem{Schroedinger} E. Schr\"odinger, {\it What is life?}, Cambridge University Press (1944)
\end{thebibliography}
\end{document}